*Review*

# Algorithmic Perspectives of Network Transitive Reduction Problems and their Applications to Synthesis and Analysis of Biological Networks

**Satabdi Aditya [1], Bhaskar DasGupta [1],* and Marek Karpinski [2]**

[1] Department of Computer Science, University of Illinois at Chicago, Chicago, IL 60607, USA;
E-Mail: aditya.satabdi@gmail.com

[2] Department of Computer Science, University of Bonn, Bonn 53113, Germany;
E-Mail: marek@cs.uni-bonn.de

\* Author to whom correspondence should be addressed; E-Mail: bdasgup@uic.edu;
Tel.: +1-312-355-1319; Fax: +1-312-413-0024.

**Abstract:** In this survey paper, we will present a number of core algorithmic questions concerning several transitive reduction problems on network that have applications in network synthesis and analysis involving cellular processes. Our starting point will be the so-called minimum equivalent digraph problem, a classic computational problem in combinatorial algorithms. We will subsequently consider a few non-trivial extensions or generalizations of this problem motivated by applications in systems biology. We will then discuss the applications of these algorithmic methodologies in the context of three major biological research questions: synthesizing and simplifying signal transduction networks, analyzing disease networks, and measuring redundancy of biological networks.

## 1. Introduction

In this survey paper, we review several transitive reduction problems on network that have applications in network synthesis and analysis involving cellular processes. Investigations of problems of these types that involve dealing with formal frameworks of very similar combinatorial nature have been done two by independent groups of communities of researchers, one being the theoretical computer science and computer networking research community and the other being the biological network research community. However, from the published literature it follows that there is minimal cooperation between such groups. The purpose of this survey is to promote a constructive dialogue among these two research communities working on similar problems so that intrigued biologists may probe further and learn new techniques from the perspective of formal analysis of algorithms and intrigued computer scientists may probe further to learn new terminologies and applications in biology. Following the general guidelines of this special issue, we first present the formal algorithmic ideas separately from their application and subsequently discuss the applications that involve these formal frameworks.

*Minimum equivalent digraph* is a classical computational problem (cf. [1]) with several recent extensions motivated by applications in social sciences and systems biology. A formal definition of the *basic* equivalent digraph problem is as follows.

**Problem name:** Minimum equivalent digraph (MIN-ED)

**Input:** a directed graph (digraph) G = (V, E).

**Definition:** for a digraph (V, E) the transitive closure of E is the relation $\stackrel{E}{\Rightarrow}$ on V × V defined as

$$u_i \stackrel{E}{\Rightarrow} u_j \equiv E \text{ contains a path from } u_i \text{ to } u_j$$

**Valid solution:** A ⊆ E such that $\stackrel{E}{\Rightarrow}$ is equal to $\stackrel{A}{\Rightarrow}$.

**Objective:** *minimize* |A|.

A complementary problem is the MAX-ED problem whose objective is to maximize |E\A|. Even though the complexity of finding an exact solution is the same for both MIN-ED and MAX-ED, the same may not necessarily be true for their approximate solutions (in the same manner as for node cover and independent set problems for general graphs [2]). For example, suppose that we have a graph with 1,000 edges and an exact solution for MIN-ED and MAX-ED with 490 edges. Suppose that an approximation algorithm for MIN-ED guarantees that we will find a solution with at most 980 edges. Thus, this approximation algorithm provides an approximation ratio of 980/490 = 2 for MIN-ED. However, the same algorithm for MAX-ED can have an approximation ratio as large as

$$\frac{1000-490}{1000-980} = \frac{510}{20} = 25.5 \tag{1}$$

Skipping the condition A ⊆ E in the definition of MIN-ED (or MAX-ED) yields the so-called transitive reduction (TR) problem which was solved in polynomial-time by Aho, Garey and Ullman [3]. See Figure 1 for an illustration of valid solutions of MIN-ED.

*1.1. Three Extensions of the Basic Version*

In this subsection, we discuss three non-trivial extensions of the basic problem that have been formulated based on their applications. We will review in more details the applications of the basic version as well as the other extensions separately in Section 4.

**Figure 1.** Illustrations of two valid solutions of MIN-ED on an input graph: (**a**) The original graph G = (V, E); (**b**, **c**) Two valid solutions (V, A$_1$) and (V, A$_2$) of MIN-ED for G. The solution in (**c**) is optimal since it has fewer edges.

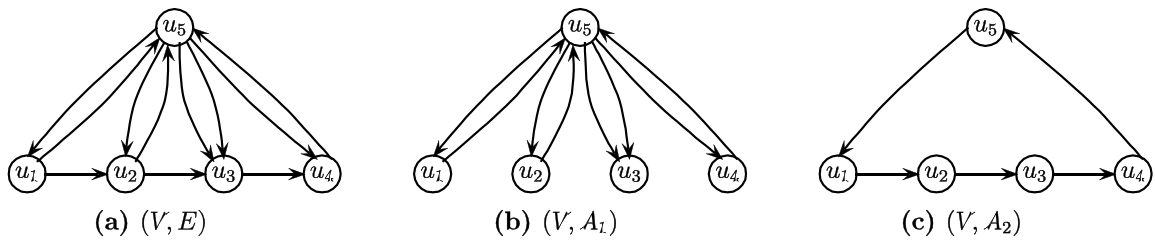

(**a**) (*V, E*)     (**b**) (*V, A$_1$*)     (**c**) (*V, A$_2$*)

1.1.1. MIN-ED and MAX-ED with Critical Edges

This extension is the same as MIN-ED or MAX-ED except that a given subset D of edges *must be present in any valid solution*. Formally, we are given D ⊆ E as part of input and the condition "A ⊆ E" is changed to D ⊆ A ⊆ E. Let us denote this version as critical-MIN-ED and critical-MAX-ED, as appropriate. As we will see subsequently, this extension is quite non-trivial if one desires a good approximate solution.

1.1.2. Weighted Version of MIN-ED or MAX-ED

In this version, each edge has a weight (positive real number) and an optimal valid solution must have the minimum possible value of total edge weights. Formally, we have a weight function $w: E \to \Re^+$ and the goal is either to minimize $\Sigma_{e \in A} w(e)$ or to maximize $\Sigma_{e \in E} w(e) - \Sigma_{e \in A} w(e)$. Let us denote this version as weighted-MIN-ED or weighted-MAX-ED, as appropriate. Obviously, the basic version is a special case of this weighted version when every edge weight is 1.

1.1.3. Binary Transitive Reduction (BTR)

This extension is a generalization of the basic version with critical edges and is described as follows [4–7]. We have an edge-labeling function $\ell: E \to \{-1, 1\}$. The label or *parity* of a path $P = (u_0, u_1, \ldots, u_k)$ is derived from the labels of its edges and given by $\ell(P) = \Pi_i \ell(u_{i-1}, u_i)$. The transitive closure relation is now generalized as $\stackrel{\ell(E)}{\Rightarrow} = \{(u_i, u_j, q) : \exists \text{ path } P \text{ using edges in } E \text{ from } u_i \text{ to } u_j \text{ and } \ell(P) = q\}$. Then, A is a binary transitive reduction of E with a required subset D if $D \subseteq A \subseteq E$ and $\stackrel{\ell(A)}{\Rightarrow} = \stackrel{\ell(E)}{\Rightarrow}$. Obviously, the basic version with critical edges is a special case of BTR when every edge label is 1. There are two (maximization and minimization) objective functions corresponding to the two generalizations of the basic version MIN-ED and MAX-ED; they will be denoted by MIN-BTR and MAX-BTR, respectively. We will use the notation $u_i \stackrel{p, E}{\Rightarrow} u_j$ to indicate a path from node $u_i$ to node $u_j$ of parity $p \in \{-1, 1\}$.

The relationships between various versions of the basic equivalent digraph problem are as follows:

$$\text{MIN-ED} < \text{Weighted-MIN-ED}$$

$$\text{MAX-ED} < \text{Weighted-MAX-ED}$$

$$\text{MIN-ED} < \text{critical-MIN-ED} < \text{MIN-BTR}$$

$$\text{MAX-ED} < \text{critical-MAX-ED} < \text{MAX-BTR}$$

where A < B means problem A is a special case of problem B. The relationships between the problem Weighted-MIN-ED and the problems critical-MIN-ED and MIN-BTR (and, similarly between the problem Weighted-MAX-ED and the problems critical-MAX-ED and MAX-BTR) are not completely known, though it is possible to design approximation algorithms for critical-MIN-ED and MIN-BTR based on approximation algorithms for Weighted-MIN-ED.

We review the following standard definitions in approximation algorithms theory. A *ε-approximate solution* (or simply a *ε-approximation*) of a minimization (respectively, maximization) problem is a polynomial-time solution with an objective value no smaller than (respectively, no larger than) ε times the value of the optimum; an algorithm of *performance* or *approximation ratio* ε produces an ε-approximate solution. A problem is *APX-hard* if there exists a ε > 1 such that *no* polynomial-time algorithm has an approximation ratio of ε *unless* P = NP. The notation OPT(G) (or simply OPT when G is clear from the context) will always denote the objective value of an optimal solution for the problem under consideration. We assume that the reader is familiar with the *basic concepts of design and analysis of algorithms* found in graduate level algorithms textbooks such as [2,8], and *basic concepts of computational biology* found in standard textbooks such as [9,10].

## 2. Summary of Known Algorithmic and Inapproximability Results

In this section, we briefly review known algorithmic and inapproximability results for the various equivalent digraph and transitive reduction problems defined in the previous section, leaving a more detailed description of algorithmic techniques used to obtain these results in the next section.

The algorithmic research work on MIN-ED was initiated by Moyles and Thomson [1] who described an efficient polynomial-time reduction of this problem for an arbitrary graph to that for a strongly connected graph, followed by an exact but exponential time algorithm for strongly connected graphs. Subsequently, an approximation algorithm for MIN-ED was detailed by Khuller, Raghavachari and Young [11] with an approximation ratio of $\left(\frac{\pi^2}{6} - \frac{1}{36} + \varepsilon\right) \approx 1.617 + \varepsilon$ (for any constant $\varepsilon > 0$), which was improved to an approximation algorithm with an approximation ratio of 3⁄2 independently by Vetta [12] and by Berman, DasGupta and Karpinski [13]. Except [13], none of these approximation algorithms will generalize directly to critical-MIN-ED with the *same* approximation ratio. The *only* non-trivial approximation algorithm known for either MAX-ED or critical-MAX-ED is a 2-approximation algorithm described in [13].

For weighted-MIN-ED, Frederickson and JàJà [14] designed a 2-approximation algorithm using an algorithm for *minimum cost rooted arborescence* due to Edmonds [15] and Karp [16]. Basically, it *suffices* to find a minimum cost in- arborescence and out-arborescence in respect to an arbitrary root node v ∈ V and take the union of all the edges in these two arborescences as the approximate solution.

Albert *et al.* [4] showed how to convert any algorithm for MIN-ED with an approximation ratio ρ to an algorithm for critical-MIN-ED with an approximation ratio of $3 - \frac{2}{\rho}$. They also provided a 2-approximation for MIN-BTR, but in fact, minor modification of their method and analysis as outlined in [13] yields a $\frac{5}{3}$-approximation. Other heuristics for these problems were investigated in [5,6] but none of these heuristics guarantees a better approximation ratio. Table 1 shows a theoretical comparison of running times and approximation ratios of some of the known algorithms for the transitive reduction problems. Unfortunately, a systematic comparative *empirical* evaluation of these algorithmic approaches is not available in the published literature. However, implementations of several algorithmic approaches on an individual level are available. For example, Kachalo *et al.* [6] provided a software called NET-SYNTHESIS which used some of the algorithmic approaches described in Sections 3.2 and 3.4, and Milanovíc *et al.* [17] discussed two meta-heuristic approaches to solve a more general version of the MIN-BTR problem.

On the inapproximability side, Papadimitriou [18] left it as an exercise to show that MIN-ED is NP-hard. Subsequently, Khuller, Raghavachari and Young [11] provided a formal proof of both NP-hardness and APX-hardness of MIN-ED for arbitrary graphs. Motivated by their cycle contraction method in [11], they were interested in the complexity of the problem when there is an *upper bound* $\gamma$ on the length of any cycle in the input graph. In [18] the authors showed that MIN-ED can be solved in polynomial time if $\gamma = 3$, MIN-ED is NP-hard if $\gamma = 5$, and MIN-ED is APX-hard if $\gamma \geq 17$. Reference [13] improved the APX-hardness result to show that both MIN-ED and MAX-ED are APX-hard even when $\gamma \geq 5$. The exact complexity of both MIN-ED and MAX-ED when $\gamma = 4$ is *still* unresolved.

**Table 1.** Theoretical comparison of worst-case performance of some of the algorithms for the transitive reduction problems.

| Problem name | Algorithmic approach | Worst-case running time using straightforward implementation | Approximation ratio |
|---|---|---|---|
| MIN-ED | Khuller, Raghavachari and Young [11] | $O(n^{1/\varepsilon})$ | $1.617 + \varepsilon^2$ |
| MIN-ED | Vetta [12] Berman, DasGupta and Karpinski [13] | $O(n \log n)$ | $\frac{3}{2}$ |
| MAX-ED | Berman, DasGupta and Karpinski [13] | $O(n \log n)$ | 2 |
| critical-MIN-ED | Khuller, Raghavachari and Young [11] | $O(n^{1/\varepsilon})$ | $2.617 + \varepsilon^2$ |
| critical-MIN-ED | Berman, DasGupta and Karpinski [13] | $O(n \log n)$ | $\frac{3}{2}$ |
| critical-MIN-ED | Frederickson and JàJà [14] | $O(n)$ | 2 |
| critical-MIN-ED | Albert *et al.* [4] | $O(n^3)$ | $\frac{5}{3}$ |
| critical-MAX-ED | Berman, DasGupta and Karpinski [13] | $O(n \log n)$ | 2 |
| weighted-MIN-ED | Frederickson and JàJà [14] | $O(n)$ | 2 |
| MIN-BTR | Albert *et al.* [4] | $O(n^3)$ | 2 |
| MIN-BTR | Berman, DasGupta and Karpinski [13] | $O(n \log n)$ | $\frac{3}{2}$ |
| MAX-BTR | Berman, DasGupta and Karpinski [13] | $O(n \log n)$ | 2 |

## 3. Review of a Few Algorithmic Techniques Used for Transitive Reduction Problems

In this section, we review a few key algorithmic techniques that have been used in the literature to investigate algorithmic complexities of various versions of the transitive reduction problem. Our goal is not to provide every technical detail involving these methods, but rather to bring our salient features of these techniques in a way that may be understood by the practitioners as well.

*3.1. From General Graphs to Strongly Connected Graphs*

Recall that a digraph (V, E) is strongly connected if and only if, for every pair of nodes $u_i$ and $u_j$, both the paths $u_i \stackrel{E}{\Rightarrow} u_j$ and $u_j \stackrel{E}{\Rightarrow} u_i$ exist. A reduction that was originally suggested in [1] and have been implicit in all subsequent works is the assumption that an ε-approximation algorithm for critical-MIN-ED and critical-MAX-ED when the given graph is strongly connected also implies an ε-approximation algorithm for the same problem on arbitrary digraphs. To understand why this is true, we first note that all these four problems can be solved easily in polynomial time using the following greedy approach if the input graph G = (V, E) is a directed acyclic graph (DAG) with D ⊆ E as the set of required edges (ϕ is the standard mathematical symbol of an empty set):

---

Compute a topological ordering $u_1, u_2, \ldots, u_n$ of the nodes of G  (* thus, if $(u_i, u_j) \in E$ then i < j *)

E' = E ; A = ϕ

**for** *i = n, n − 1, n − 2, …, 1* **do**

    **for** *j = n, n − 1, n − 2, …, i + 1* **do**

        **if** $(u_i, u_j) \in E$ **then**

            **if** $(u_i, u_j) \in D$ **then** add the edge $(u_i, u_j)$ to A

            **else** if the path $u_i \stackrel{E \setminus \{u_i, u_j\}}{\Rightarrow} u_j$ does not exist **then** add the edge $(u_i, u_j)$ to A

***Return*** *(V, A) as the solution*

---

It is easy to implement the above algorithm to run in polynomial time. Now, suppose that the input graph G is not a DAG and consider the strong component graph G' = (V', E') of G:

$$V' = \{C \mid C \text{ is a strongly connected component of G}\}$$

$$E' = \{(C, C') \mid C.C' \in V' \text{ and } (u_i, u_j) \in E \text{ for some } u_i \in C \text{ and } u_j \in C'\}$$

It is easy to see that G' is a DAG and can be found in O(|V| + |E|) time [8]. Let A' be the solution of our problem on G'. Suppose that we have ε-approximation algorithm for critical-MIN-ED or critical-MAX-ED on each strongly connected component of G. Then, the union of the edges in this ε-approximation for every strongly connected component of G together with the edges in A' provide an ε-approximation for the entire graph G.

For MIN-BTR or MAX-BTR Albert *et al.* [4] provides a more complex reduction to show that an ε-approximation algorithm for strongly connected graphs also implies an ε-approximation algorithm for arbitrary digraphs. To achieve this, each strongly connected component is replaced a graph with constantly many edges and nodes (called "gadget" in [4]) and then these graphs are connected appropriately such that the resulting graph is a DAG and an ε-approximation for the entire graph can be recovered using an exact optimal solution of the DAG and ε-approximations of the strongly connected components.

*Thus*, *for the remainder of this section*, we assume without loss of generality that the input graph G is strongly connected.

*3.2. The Cycle Contraction Method [11]*

Consider an input graph G = (V, E) for the MIN-ED problem and suppose that G has a directed Hamiltonian cycle, *i.e.*, a (directed) cycle that contains every node exactly once. Then clearly the edges in this cycle constitute an optimal solution of |V| edges. This intuition suggests a general strategy of repeatedly finding a longest cycle in the given graph, selecting the edges in this cycle and modifying the graph to reflect the selection of edges until we reach a valid solution.

However, finding a directed Hamiltonian cycle or the longest cycle is in general NP-hard [2]. To circumvent the NP-hardness issue, Khuller, Raghavachari and Young in [11] designed the following "cycle contraction" approach. Contraction of an edge ($v_i$, $v_j$) is nothing but the act of merging the two nodes $v_i$ and $v_j$ into a new single node $v_{ij}$ and deleting any resulting self-loops or multi-edges. Similarly, contraction of a cycle is defined as the contraction of every edge of the cycle; see Figure 2 for an illustration. Note that if c is a constant then one can easily check in polynomial time if a graph has a cycle of at least c edges. The algorithm, parameterized by a constant c > 3 to be chosen by the user, now proceeds as follows:

> **for** $i = c, c - 1, \ldots, 4$ **do**
>
>     **while** (the graph contains a cycle of at least *i* edges) **do**
>
>         Find a cycle C of at least *i* edges
>
>         Select the edges in C and contract C
>
>     **endwhile**
>
> **endfor**
>
> *(\* now the graph contains no cycle of more than 3 edges \*)*
>
> Solve MIN-ED on the reduced graph exactly using the algorithm in [19] and select the edges in this exact solution.

**Figure 2.** Illustration of a cycle contraction: (**a**) shows the original graph and (**b**) shows the graph after the cycle $u_1, u_2, u_3, u_4, u_5, u_6, u_1$ has been contracted.

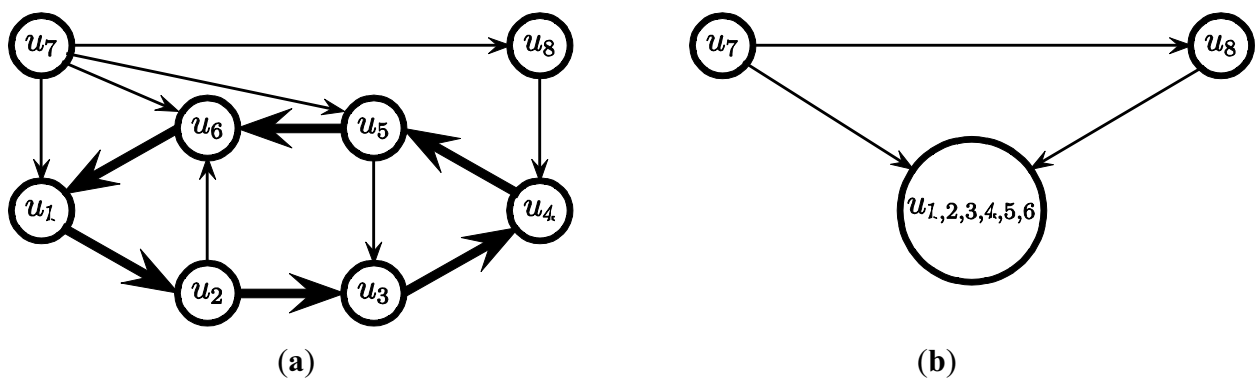

(**a**)          (**b**)

It was shown in [11] that the above algorithm for MIN-ED returns a valid solution containing y edges where $y \leq \left( \dfrac{\pi^2}{6} - \dfrac{1}{36} + \dfrac{1}{c(c-1)} \right) \text{OPT}(G) \approx \left( 1.617 + \dfrac{1}{c(c-1)} \right) \text{OPT}(G)$ edges.

The above approach can also be applied to critical-MIN-ED by simply adding all the edges from the required set of edges D to the solution. The number of edges z in the resulting solution of critical-MIN-ED satisfies $z \leq \left(1 + \frac{\pi^2}{6} - \frac{1}{36} + \frac{1}{c(c-1)}\right) \text{OPT}(G) \approx \left(2.617 + \frac{1}{c(c-1)}\right) \text{OPT}(G)$ since obviously |D| ≤ OPT. Another possibility outlined in [4] is to replace every required edge $(u_i, u_j) \in D$ by introducing a new node $u_{ij}$ and adding two new edges $(u_i, u_{ij})$ and $(u_{ij}, u_j)$, running the approximation algorithm for MIN-ED on this new graph, and then replacing the edges $(u_i, u_{ij})$ and $(u_{ij}, u_j)$ in the solution by the original edge $(u_i, u_j)$. If an optimal solution of critical-MIN-ED on G uses β edges from E\D then this approach returns a solution (V,A) with

$$|A| \leq \left(1 + \frac{\pi^2}{6} - \frac{1}{36} + \frac{1}{c(c-1)}\right)(2|D| + \beta) - |D| \approx 2.236|D| + 1.618\beta.$$

*3.3. The Arborescence Approach [14]*

A (rooted) *spanning out-arborescence* of a directed edge-weighted graph G = (V, E) is a directed acyclic spanning sub-graph (V, A) of G such that every node except one node (the *root*) has *exactly* one incoming edge and the weight of such an out-arborescence is the sum of the weight of its edges. A spanning *in-arborescence* is defined analogously except that every node except the root has exactly one *outgoing* edge. An exact polynomial-time solution for computing a spanning in-arborescence or spanning out-arborescence of minimum weight was provided by the authors in [15,16,20]. An overview of this algorithm for computing a minimum weight out-arborescence (as formulated in [16]) is as follows. We first remove all incoming edges to the root *v*. Then we proceed as follows. First, we select for each node, except the root *v*, an incoming edge of minimum weight. If these edges do not give a spanning arborescence, then there must be a (directed) cycle C formed by a subset of these edges. Let $w(C) = \min \{w(e) | e \in C\}$. We contract the cycle C to a "mega"-node, and decrease the weight of every edge $(u, v)$ from a node $u \notin C$ to a node $v \in C$ by α − w(C), where α is the weight of the *unique* edge in C that is incoming to *v*. The process is then *repeated* on the reduced graph, and continued until we have a spanning arborescence on the remaining graph. The mega-nodes are then expanded in the *reverse* order. Each time a mega-node is expanded, *exactly* one of its edges that would produce two incoming edges to a node is discarded. A minimum weight in-arborescence can be computed by the *same* algorithm if we *reverse* the direction of all the edges of the input graph. See Figure 3 for an illustration.

For weighted-MIN-ED, Frederickson and JàJà [14] proposed the following simple algorithm that gives a 2-approximation for an input graph G = (V, E):

    Select an arbitrary node v of G
    Find a minimum weight spanning in-arborescence (V, $A_1$) of G rooted at *v*
    Find a minimum weight spanning out-arborescence (V, $A_2$) of G rooted at *v*
    Return (V, $A_1 \cup A_2$) as the solution

**Figure 3.** An illustration of the algorithm to compute a minimum weight spanning out-arborescence. The thick black edges at the final fourth step are the edges in the solution.

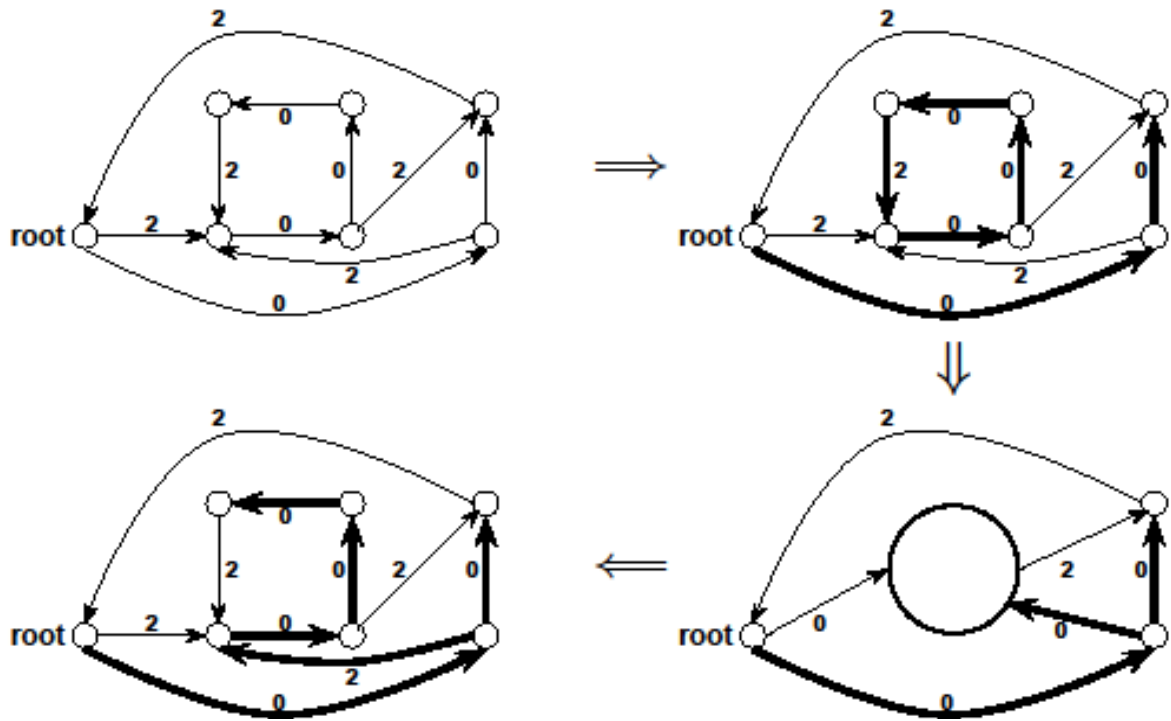

The above solution is a *valid* solution since we can reach any node $v_j$ starting from any node $v_i$ by taking a path from $v_i$ to the root $v$ followed by a path from $v$ to the node $v_j$. The solution is a 2-approximation since any valid solution of weighted-MIN-ED includes both a minimum weight spanning in-arborescence and a minimum weight spanning out-arborescence and thus $OPT(G) \geq \max\{|A_1|, |A_2|\}$. A simple example of an input graph was also provided in [14] for which the above algorithm provides a solution to total weight $2OPT(G)$.

For critical-MIN-ED, a very similar approach as described below can be used to again provide a 2-approximation for an input graph $G = (V, E)$:

Define the weight *w(e)* of an edge $e \in E$ as $w(e) = \begin{cases} 0, & \text{if } e \in D \\ 1, & \text{otherwise} \end{cases}$

Select an arbitrary node $v_r$ of G

Find a minimum weight spanning in-arborescence $T = (V, A_1)$ of G rooted at node $v_r$

Redefine the weight *w(e)* of an edge $e \in E$ as $w(e) = \begin{cases} 0, & \text{if } e \in D \cup A_1 \\ 1, & \text{otherwise} \end{cases}$

Find a minimum weight spanning out-arborescence $T = (V, A_2)$ of G rooted at node $v_r$

Return $(V, A_1 \cup A_2 \cup D)$ as the solution

Albert *et al*. [4] showed how to modify the above algorithm and combine it with any ρ-approximation algorithm for MIN-ED to obtain an *improved* algorithm for critical-MIN-ED with an approximation ratio of $3 - \frac{2}{\rho}$. Currently, the best possible value of ρ is 1.5 which leads to a ⁵⁄₃-approximation for critical-MIN-ED using this approach.

*3.4. From Critical-MIN-ED And Critical-MAX-ED To MIN-ED And MAX-ED [4,13]*

The results in [4,13] show how to transform a solution to critical-MIN-ED (respectively, critical-MAX-ED) to a solution to MIN-ED (respectively, MAX-ED) by adding a single edge (We remind the reader that we assume that the input graph is strongly connected.) that can be found in polynomial time. The idea behind this is as follows. We can distinguish our input (and strongly connected) graph G based on whether G = (V, E) has a cycle of parity −1 (double parity graph) or not (single parity graph). Whether G is a single or double parity graph can be easily checked in $O(|V|^3)$ time by using a simple modification of the well-known Floyd-Warshall transitive closure algorithm [8] as outlined in [4]. Now we can observe the following:

- If G is a single parity graph then for every pair of nodes $u_i, u_j \in V$, exactly one of the two the paths $u_i \stackrel{1,E}{\Rightarrow} u_j$ and $u_i \stackrel{-1,E}{\Rightarrow} u_j$ exists. Then, we can simply ignore the edge labels and compute a solution (V, A) of critical-MIN-ED (respectively, critical-MAX-ED) on G. It can be seen that (V, A) also provides a valid solution for MIN-ED (respectively, MAX-ED).
- Otherwise, G is a double parity graph. We again first ignore the edge labels and compute a solution (V, A) of critical-MIN-ED (respectively, critical-MAX-ED) on G. Note that (V, A) contains a rooted arborescence, say (V, $A_1$) with $A_1 \subseteq A$, rooted at some node $u_r$. We label each node $u_i \in V$ with $\ell(u_i) = \ell(P_i)$ where $P_i$ is the *unique* path in (V, $A_1$) from $u_r$ to $u_i$. Since G is a double parity graph, there must exist an edge $(u_i, u_j) \in E$ such that $\ell(u_i) \ell(u_j) \neq \ell(u_i, u_j)$, and adding this edge (if not already present) to A produces a valid solution of critical-MIN-ED or critical-MAX-ED for G.

*3.5. Linear Programming Based Approach [13]*

We refer the reader to a standard graduate level textbook such as [21] for basic concepts and definitions related to linear programming and its applications to designing approximation algorithms.

An exponential-size linear programming (LP) formulation for the minimum weight rooted (at node $u_r$) out-arborescence problem for an edge-weighted input graph G = (V, E) was provided by Edmonds [15] in the following manner. We use a *binary* indicator variable $x_e = x_{u_i, u_j}$ for every edge $e = (u_i, u_j) \in E$ which describes whether we select $e$ ($x_e = 1$) or do not select $e$ ($x_e = 0$) in our solution. For $U \subset V$, define $\iota(U) = \{(u_i, u_j) \in E : u_i \notin U \text{ and } u_j \in U\}$. Then, the LP formulation is:

$$\begin{aligned}
&\text{minimize} \sum_{e \in E} w(e) x_e \\
&\text{subject to} \\
&\sum_{e \in \iota(U)} x_e \geq 1 \text{ for all U such that } \Phi \subset U \subset V \text{ and } u_r \notin U \\
&x_e \geq 0 \text{ for all } e \in E
\end{aligned} \quad (2)$$

Edmonds [15] showed that the above LP always has an integral optimal solution (*i.e.*, an optimal solution with $x_e \in \{0, 1\}$ for all $e \in E$) which provides an optimal solution for our minimum weight rooted out-arborescence problem. Note that the above LP has $O(2^{|V|})$ constraints in the worst case.

However, the advantage of such a linear programming is that we can now make use of powerful mathematical tools, such as the duality theorem, from the theory of linear programming.

We can modify the above LP formulation to a *primal* LP formulation $P_1$ for MIN-ED provided we set $w(e) = 1$ for all $e \in E$ and we remove "and $u_r \notin U$" from the condition in constraint (1). The *dual* program $D_1$ of this LP can be constructed by having a variable $y_U$ for *every* $\Phi \subset U \subset V$. Both the primal and the dual LP are written down below for clarity.

| (primal LP $P_1$) | (dual LP $D_1$) |
|---|---|
| minimize $\sum_{e \in E} x_e$ | maximize $\sum_{\Phi \subset U \subset V} y_U$ |
| subject to | subject to |
| $\sum_{e \in \iota(U)} x_e \geq 1$ for all U such that $\Phi \subset U \subset V$ | $\sum_{\substack{\Phi \subset U \subset V \\ e \in \iota(U)}} y_U \leq 1$ for every edge $e \in E$ |
| $x_e \geq 0$ for all $e \in E$ | $y_U \geq 0$ for all $\Phi \subset U \subset V$ |

We can change $P_1$ into a LP formulation for MAX-ED if we replace the objective "minimize $\Sigma_{e \in E} x_e$" by "maximize $|E| - \Sigma_{e \in E} x_e$", and change the dual $D_1$ accordingly to reflect this change. We can further change this formulation for MIN-ED and MAX-ED to critical-MIN-ED and critical-MAX-ED, respectively, by adding a constraint $x_e \geq 1$ for every edge $e \in D$.

Note that $P_1$ does not provide a valid solution of the MIN-ED problem unless the constraint $x_e \geq 0$ for every edge $e \in E$ is replaced by the constraint $x_e \in \{0, 1\}$, resulting in an integer linear program (ILP) whose exact solution is in general NP-hard to compute. We will denote this ILP corresponding to $P_1$ by $IP_1$.

3.5.1. Applying LP-Based Approach to Critical-MIN-ED

We provide a *high-level* overview of the primal-dual approach used in [13] for critical-MIN-ED on an input graph $G = (V, E)$.

1. We start with an initial assignment of values to variables in **IP$_1$** in the following manner. We keep only a subset of constraints of **IP$_1$** such that the resulting ILP can be solved *exactly* in *polynomial* time, giving an optimal solution $A_1 \subseteq E$. Then, it follows that $OPT(G) \geq |A_1|$.
2. However, $(V, A_1)$ may not be a valid solution for critical-MIN-ED on G (*i.e.*, **IP$_1$**). Then, we try to make $A_1$ a valid solution by adding and/or removing edges so that we use a total of at most $\frac{3}{2}\eta - 1$ edges where $OPT(G) \geq \eta \geq |A_1|$, giving a ³⁄₂ -approximation for critical-MIN-ED. The edge alteration procedure was carried out in [13] using the DFS (depth-first-search) algorithm as originally outlined in a seminal paper by Tarjan (e.g., see the textbook [22]).

The initial solution $A_1$ referred to above in Step 1 is obtained in the following manner. For $U \subset V$, define $o(U) = \{(u_i, u_j) \in E : u_i \in U \text{ and } u_j \notin U\}$. Call a constraint of type $\Sigma_{e \in \iota(U)} x_e \geq 1$ in IP$_1$ "tractable" if for some node $u_i$ either $\iota(U) \subseteq \iota(\{u_i\})$ or $\iota(U) \subseteq o(\{u_i\})$. It was shown in [13] that the set of tractable constraints of IP$_1$ can be found easily and the resulting ILP can be solved exactly using any algorithm that finds a *maximum matching in a bipartite graph*. Figure 4 shows an example of the initial solution $A_1$ found by this approach.

**Figure 4.** An illustration of the initial solution $A_1$ discussed in the algorithm that applies a LP-based approach to critical-MIN-ED in Section 3.5.1: (**a**) The input graph G; (**b**) The edges in the initial solution $A_1$. As one can see, the initial solution does *not* provide a valid solution of critical-MIN-ED since the graph in Figure 4b is not strongly connected, but the final solution is obtained by adding and deleting edges from this initial solution.

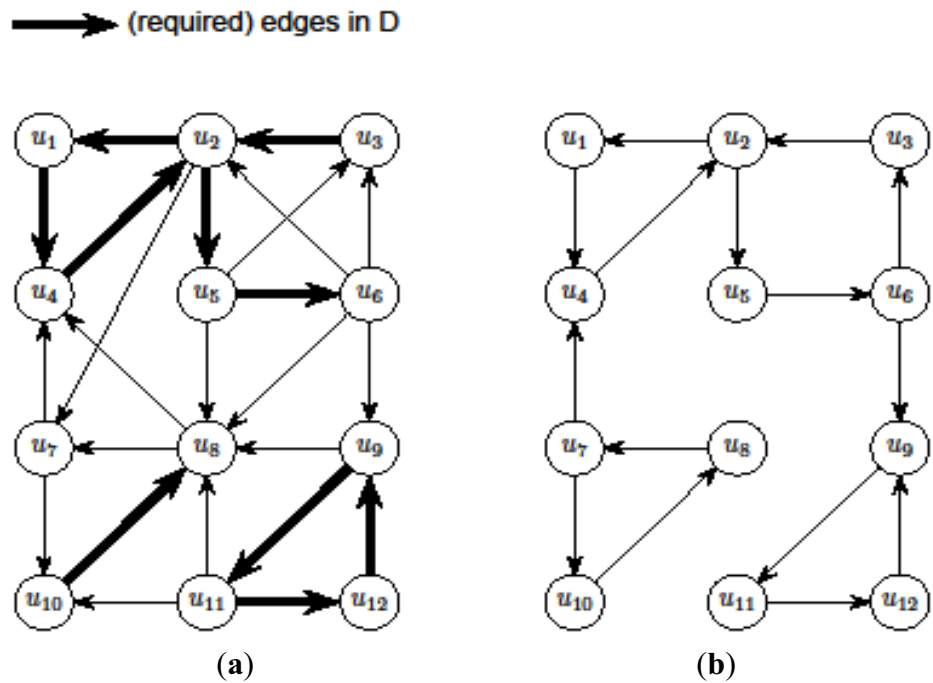

(**a**)            (**b**)

The DFS-based edge addition/removal method referred to in Step 2 is *highly* technical with elaborate case analysis and is beyond the scope of this review paper. In a nutshell, difficulties may arise because in some cases the algorithm may be *forced* to use more than $\frac{3}{2}|A_1|-1$ edges. Then, we look at the "non-tractable" constraints of the primal $P_1$ or dual $D_1$ to get an improved lower-bound $\eta$ for OPT(G) (*i.e.*, OPT(G) $\geq \eta > A_1$) to ensure that we use at most $\frac{3}{2}\eta - 1$ edges. In the proof we need to crucially use the weak-duality theorem of linear programming which states that if OPT($P_1$) and OPT($D_1$) are the objective values of an optimal solution of $P_1$ and $D_1$, respectively, then OPT($P_1$) $\geq$ OPT($D_1$).

3.5.2. Applying LP-Based Approach to Critical-MAX-ED

We provide an overview of the 2-approximation algorithm for critical-MAX-ED on an input graph G = (V, E) using a LP-based approach as described in [13]. Call an edge $e \in$ E a *necessary* edge if either e $\in$ D or $\iota(U) = \{e\}$ for some U $\subset$ V and let F be the set of necessary edges. If the edges in F provide a valid solution of critical-MAX-ED on G then (V, F) provide us with an optimal solution, thus assume that this is not the case below. In this case, $\sum_{e \in \iota(U)} = 0$ for some $\Phi \subset$ U $\subset$ V, so there must be a node $u_r$ such that no edges in F enter $u_r$. As a *pre-processing* step, we repeatedly contract a cycle of necessary edges until no such cycles remain. Let OPT$_{\text{in-arb}}$(G) be the total weight of a minimum-weight in-arborescence of G rooted at $u_r$. Consider the LP formulation for the minimum weight rooted out-arborescence problem as defined before:

$$\text{minimize} \sum_{e \in E} w(e) x_e$$

$$\text{subject to}$$

$$\sum_{e \in \iota(U)} x_e \geq 1 \text{ for all U such that } \Phi \subset U \subset V \text{ and } u_r \notin U$$

$$x_e \geq 0 \text{ for all } e \in E$$

and let $w(e) = \begin{cases} 0, \text{ if } e \in F \\ 1, \text{ otherwise} \end{cases}$. Now, suppose that we set $x_e = \begin{cases} 1, \text{ if } e \in F \\ \frac{1}{2}, \text{ otherwise} \end{cases}$. This assignment of variables is a *valid* solution of the above LP.

Now, compute a minimum weight out-arborescence $T_{out} = (V, A_{out})$ rooted at $u_r$. If there are $z + 1$ edges in E that are *not* in $A_{out}$, then $OPT(G) \leq z$. Suppose now that we change $w(e)$ for every $e \in A_{out}$ to zero and keep the other weights *unchanged*. Our previous fractional solution, namely $x_e = \begin{cases} 1, \text{ if } e \in F \\ \frac{1}{2}, \text{ otherwise} \end{cases}$, is still a valid solution of the LP, and thus the total value of the objective function of this fractional solution is at most $\frac{z+1}{2}$, which together with the result of Edmonds [15] that showed that "the LP always has an integral optimal solution" implies that $OPT_{\text{in-arb}}(G) \leq \frac{z+1}{2}$, which implies that we delete at least $z + 1 - \frac{z+1}{2} = \frac{z+1}{2}$ edges from the in-arborescence and take the remaining edges of the in-arborescence together with all the edges in $A_{out}$ to get a valid solution of critical-MAX-ED on G. The total number of edges we have deleted in at least $\frac{z+1}{2} \geq \frac{OPT(G)+1}{2}$. A slight modification in the argument shows that in fact we can delete at least $\frac{OPT(G)}{2}$ edges.

### 3.5.3. Limitations of LP-Based Approaches

A standard way of understanding the limitations of any LP-based approach for designing approximation algorithms is to measure the *integrality gap*, *i.e.*, the ratio of the objective value of an optimal integral solution to that of an optimal fractional solution for a minimization problem and the ratio of the objective value of an optimal fractional solution to that of an optimal integral solution for a minimization problem [21]. In [13] it was shown that the integrality gap for $P_1$ was at least $\frac{4}{3}$ by giving an explicit construction of an input graph for which this ratio is achieved. The same input graph also shows that the integrality gap for the modification of $P_1$ corresponding to MAX-ED is at least $\frac{3}{2}$.

## 4. Biological Applications

In this section, we discuss three applications of transitive reduction problems in computational biology and bioinformatics. For other non-biology applications of transitive reduction problems, such as in *visualization* of Enron email networks or in *connectivity* issues of *computer* networks, the reader may consult appropriate references such as [11,23].

We briefly review the standard regulatory network model that was mentioned in Section 1.1.3 in connection with the MIN-BTR and MAX-BTR problems. A regulatory network is described by an *edge-labelled directed* graph G = (V, E) in which nodes represent individual components of the biological system and (directed) edges of the form $(u_i, u_j)$ indicates that node $u_i$ has an influence on node $u_j$. The edge labelling function $\ell: E \to \{-1, 1\}$ indicates the *nature* of the causal relationship, with $\ell(u, u_j) = 1$ and $\ell(u_i, u_j) = -1$ indicating that $u_i$ has an *excitatory* (positive) and *inhibitory* (negative) influence on $u_j$, respectively; pictorially, it is quite common to denote an excitory and an inhibitory edge by $\rightarrow$ and $\dashv$, respectively. This representation applies to both gene regulatory networks (describing the regulation of gene transcription and related processes) and signal transduction networks (describing the information flow from external signals to within-cell components). Some examples of large size biological networks include:

- Mammalian network of signaling pathways and cellular machines in the hippocampal CA1 neuron having 512 nodes and 1,047 edges [24].
- *S. cerevisiae* transcriptional regulatory network of interactions between transcription factor proteins and genes having 690 nodes and 1,082 edges [25].
- *C. elegans* metabolic network having 651 nodes and 2,040 edges [26].
- Oriented version of an unweighted PPI network constructed from *S. cerevisiae* interactions in the BioGRID database having 786 nodes and 2,453 edges [27].

Existence of such large networks rules out exact brute-force calculations of optimal solutions of transitive reduction problems and provides motivations to explore approximation algorithms for these problems.

*4.1. Network Construction and Simplification from Direct and Double-Causal Data*

Signal transduction and gene regulatory networks are crucial to the maintenance of cellular *homeostasis* and for cell behavior such as growth, survival, *apoptosis*, and movement. Deregulation of these networks is a key contributor to many *disease* processes such as developmental disorders, diabetes, vascular diseases, and cancer. In a signal transduction network (*pathway*), there is typically an input, perceived by a *receptor*, followed by a series of elements through which the signal percolates to the output node, which represents the final outcome of the signal transduction process. For a cellular signal transduction pathway not involving alterations in gene expression, elements often consist of proteinaceous receptors, intermediary signaling proteins and metabolites, effector proteins, and a final output, which represents the ultimate combined effect of the effector proteins. If the signal transduction process includes regulation of the transcript level of a particular gene, the intermediate signaling elements will also include the gene itself and the transcription factors that regulate it, as well any small RNAs that regulate the transcript's abundance, with the final output being presence or absence of transcripts. Genome-wide experimental methods now identify interactions among thousands of proteins [28–34]. However, the state of the art understanding of many signaling processes is *often* limited to the knowledge of key mediators and of their positive or negative effects on the whole process. The experimental evidence about the involvement of specific components in a given signal transduction network frequently belongs to one of these two categories:

- **(i)** "Direct" interactions corresponding to *biochemical* evidences that provide information on enzymatic activity or protein-protein interactions and represent direct *physical* interactions. An interaction of this type is of the form "A promotes B" or "A inhibits B", and is represented in the usual manner by a directed edge A $\rightarrow$ B and A $\dashv$ B, respectively. Edges corresponding to known (documented) direct interactions are marked as "critical" and belong to the set D of required edges.

(ii) "Putative" interaction patterns that arise, for example, during *differential* responses to a stimulus, which in a wild-type organism *versus* a mutant organism implicates the product of the mutated gene in the signal transduction process. This type of interaction pattern is not a direct interaction but rather corresponds to an *indirect* (*double-causal*) relationship most likely resulting from a *chain* of direct interactions and reactions, and is a 3-component inference represented by a small-size sub-graph among three or four nodes.

As noted above, inference of type (ii) may not give direct interactions but indirect causal relationships that correspond to reachability relationships in the unknown interaction network for which the MIN-BTR and MAX-BTR problems become directly applicable. More precisely, inferences of type (ii) typically lead to double-causal inferences of the type "C promotes the *process* through which A promotes B", and may correspond to an *intersection* of two paths (one path from A to B and another path from C to B) in the interaction network (*i.e.*, C is assumed to activate an *unknown* intermediary node of the A to B path).

The research works in [5–7] led to the development of an efficient and accurate method incorporating all relevant biological knowledge for synthesizing path-level information into a consistent network by constructing a minimal graph that maintains all reachability relationships without requiring expression information (unlike, say, many *reverse-engineering* approaches). Methods prior to [5–7] for synthesizing signal transduction networks, such as [28], only included direct biochemical interactions and were therefore restricted by the incompleteness of the experimental knowledge on pairwise interactions. Key steps in the network synthesis method developed in [5–7] are schematically shown in Figure 5. The first step is a *distillation* of experimental conclusions into qualitative regulatory relations between cellular components (This is a complex process by itself. It is important to note that human intervention will *inevitably* be an important component of the literature curation process even though automated text search engines such as GENIES [32] become more and more popular). Direct biochemical and pharmacological evidences, such as "A promotes B" are incorporated as a directed edge (A, B). Other kind of double-causal evidences (such as genetic evidences of differential responses to a stimulus) are handled in the third step in the schematic diagram. For the sake of concreteness, assume that such a double-causal interaction is of the form "C promotes the process through which A promotes B". The only way such a double-causal interaction may correspond to a direct interaction is if C is an enzyme *catalyzing* a reaction in which A is transformed into B, and for this case the interaction can be represented as both A (the *substrate*) and C (the enzyme) activating B (the *product*), *i.e.*, by two edges A $\to$ B and C $\to$ B. If the interaction between A and B is direct and C is *not* a catalyst of the interaction between A and B, we can assume that C activates A. In all other cases, this type of interaction corresponds to an intersection of two paths (A to B and C to B) in the interaction network by introducing new nodes (called "pseudo-nodes" in [5] and elsewhere since they are added only to satisfy the pathway properties). One important algorithmic idea in this network synthesis method is that of finding a *minimal* (Intuitively, by computing a minimal graph we want to be as close as possible to a "tree-like topology" while supporting all experimental observations. Implicit assumption of chain-like or tree-like topologies permeates the traditional molecular biology literature, e.g., signal transduction and metabolic pathways are assumed to be close to linear chains and genes are assumed to be regulated by one or two transcription factors [33].)

network, in terms of number of non-critical edges (*i.e.*, edges *not* in D), that is *consistent* with all (directed) reachability relationships between nodes, and is captured by the MIN-BTR and MAX-BTR problem discussed earlier. For further details, see [5–7]. A software named NET-SYNTHESIS incorporating the method shown in Figure 5 using some of the algorithmic ideas described for MIN-BTR and MAX-BTR in Section 3 was first reported in [5,6] and is freely available for download. The input to NET-SYNTHESIS is a list of relationships among biological components (direct and double causal), and its output is a network diagram and a text file with the edges of the signal transduction network.

**Figure 5.** A schematic diagram of the network synthesis method in [5–7]. Human interaction is necessary since some choices may have to be made in distilling the component relationships, e.g., when there are conflicting reports in the literature.

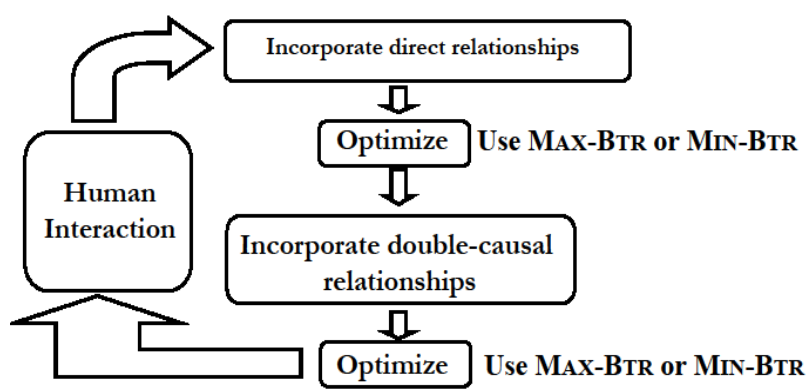

4.1.1. Applications in Agronomic Research

Guard cells are central components in control of plant water status [34] and better understanding of their regulation is imperative for the goal of engineering of crops with improved drought tolerance. Plants both lose water and take in carbon dioxide through microscopic stomatal pores, each of which is regulated by a surrounding pair of guard cells. During drought, the plant hormone *abscisic acid* (ABA) inhibits stomatal opening and promotes stomatal closure, thereby promoting water conservation. ABA signal transduction in guard cells is one of the best characterized signaling systems in plants with many signal transduction proteins, secondary metabolites and ion channels having been identified to participate in the process [35–37].

The research works in [5,6] used the NET-SYNTHESIS software to generate a network for ABA-induced closure from is a list of about 140 interactions and causal inferences for ABA-induced closure published in Table S1 and Text S1 in [38]. A detailed comparison of this computer generated network with a manually curated network for ABA-induced closure published in [38] validated the accuracy of the algorithms for MIN-BTR used in the software.

*4.2. Analyzing Disease Networks (Biomedical Application)*

Large Granular Lymphocytes (LGL) are medium to large size cells with eccentric nuclei and abundant cytoplasm. In normal adults, LGL comprise 10%~15% of the total peripheral blood mononuclear cells. The disease LGL leukemia is a disordered clonal expansion of LGL and their invasions in the marrow, spleen and liver. Ras is a small GTPase, which is essential for controlling multiple essential signaling pathways, and its deregulation is frequently seen in human cancers. Activation of H-Ras required its farnesylation, which can be blocked by farnesyltransferase inhibitiors (FTIs). This envisions FTIs as future drug target for anti-cancer therapies. One of these FTI is tipifarnib which shows apoptosis induction effect to leukemic LGL in vitro. This observation, together with the finding that Ras is constitutively activated in leukemic LGL cells, leads to the hypothesis that Ras plays an important role in LGL leukemia, and may function through influencing Fas/FasL pathway.

Kachalo *et al.* in [6] used the NET-SYNTHESIS software together with its specific transitive reduction algorithms to synthesize a cell-survival/cell-death regulation related signaling network from the Transpath 6.0 database with additional information manually curated from literature search, having 359 nodes representing proteins/protein families and mRNAs participating in pro-survival and Fas-induced apoptosis pathways and 1,295 edges representing regulatory relationships between nodes, including protein interactions, catalytic reactions, transcriptional regulation and known double-causal regulations. Using MIN-BTR and other algorithms, they were able to reduce the size of the original network to 267 nodes and 751 edges to focus special interest on the effect of Ras on apoptosis response through Fas/FasL pathway that involve the 33 known T-LGL deregulated proteins. Further work in this direction was done by Zhang *et al.* in [39] in building and analyzing a network model of signaling components of survival of cytoxic T lymphocytes in LGL-leukemia using the NET-SYNTHESIS software.

For further applications of transitive reduction problems to drug target identification, see [40].

*4.3. Measuring Topological Redundancy of Biological Networks*

The concept of redundancy is well known in information theory. Informally, redundancy refers to identical elements performing the same function (There are also other definitions of the redundancy concept in the context of other biological applications that is completely different from ours. For example, in some context redundancy refers to paralogous genes that provide functional backup for one another [41]). In computer networks and electronic systems, such measures are useful in analyzing properties such as fault-tolerance. It is an accepted fact that biological networks do *not* necessarily have the lowest possible degeneracy or redundancy. For example, the connectivity of neurons in brains suggests a high degree of degeneracy [42]. As Tononi, Sporns and Edelman observed in [43], a specific and useful notion of redundancy has yet to be firmly incorporated into biological thinking, often because of the lack of a suitable formal theoretical framework. A further reason for the lack of incorporation of these notions in biological thinking is the lack of computationally efficient procedures for computing these measures for large-scale networks even when formal definitions are available. Therefore, such studies are often done in a somewhat *ad-hoc* fashion, such as in [44]. There are notions of redundancy available in the field of analysis of undirected graphs based on *clustering coefficients* [45] or *betweenness centrality measures* [46]. However, such notions are *not* appropriate for the analysis of biological networks where we *must* distinguish positive from negative regulatory interactions or where we wish to study possible relationships of the *dynamics* of the network with its redundancy.

Based on the MIN-BTR and MAX-BTR problems, Albert *et al*. in [47] proposed a new combinatorial measure of redundancy that is amenable to efficient algorithmic analysis. Note that binary transitive reduction of a graph (V, E) does *not* change pathway level information of the network and removes an edge from one node $u_i \to u_j$ or $u_i \dashv u_j$ only when a similar *alternate* pathway, namely $u_i \overset{1, E\setminus\{u_i,u_j\}}{\Rightarrow} u_j$ or $u_i \overset{-1, E\setminus\{u_i,u_j\}}{\Rightarrow} u_j$ respectively, exists, thus truly removing redundant connections. Thus, if $(V, E_1)$ is an optimal solution of MIN-BTR and MAX-BTR on the input graph G = (V, E) then $\frac{|E_1|}{|E|}$ provides a measure of *global* compressibility of the network. Based on this intuition, Albert *et al*. in [47] proposed a new redundancy measure $R = 1 - \frac{|E_1|}{|E|}$, where the $|E|$ term in the denominator is simply a "min-max normalization" of the measure to ensure that $0 < R < 1$. Note that the higher the value of R is, the *more* redundant the network is. Since MIN-BTR or MAX-BTR can be computed efficiently, Albert *et al*. were able to evaluate R on a variety of large biological and directed social networks to derive interesting conclusions such as transcriptional networks are less redundant than signaling networks, directed social networks are more redundant than biological networks, the topological redundancy of the *C. elegans* metabolic network is largely due to its inclusion of currency metabolites and the redundancy of signaling networks is highly (*negatively*) correlated with the *monotonicity* of their dynamics.

## 5. Conclusions

In this review paper, we have elaborated on a few graph-theoretic problems that involve finding an "equivalent" sparser graph, explain several key mathematical and algorithmic tools that may be used to design efficient computational methods to solve these problems and then provided details of three biological applications of these problems. The idea of transitive reductions, in a more simplistic setting or in a different form, has also been used to identify structure of gene regulatory networks [48–52]. Of particular interest is a network "deconvolution" problem, considered by Feizi *et al*. [52], that is in some sense an inverse of the transitive reduction problems studied in this paper: their goal was to infer the original network given a set of direct (edge-level) and indirect (pathway-level) information about the graph. The authors in this paper showed that an exact closed-form solution of this problem can be found using an infinite-series summation. We hope that our review will lead to further interests in transitive reduction type problems and will promote further collaboration between the computational biology and the graph algorithms community.

## Acknowledgments

S. Aditya and B. DasGupta was partially supported by NSF grants IIS-1160995 and DBI-1062328. B. DasGupta also thanks his collaborators R. Albert, P. Berman, R. Dondi, A. Gitter, G. Gürsoy, R. Hegde, S. Kachalo, P. Pal, G. S. Sivanathan, E. D. Sontag, A. Zelikovsky, K. Wesrbrooks and R. Zhang for their collaboration in the research projects reviewed in this paper.

## Conflicts of Interest

The authors declare no conflict of interest.